# A Revision Control System for Image Editing in Collaborative Multimedia Design

Fabio Calefato, Giovanna Castellano, Veronica Rossano

University of Bari

Italy

{fabio.calefato, giovanna.castellano, veronica.rossano}@uniba.it

*Abstract*—Revision control is a vital component in collaborative development of artifacts such as software code and multimedia. While revision control has been widely deployed for text files, very few attempts to control the versioning of binary files can be found in the literature. This can be inconvenient for graphics applications that use a significant amount of binary data, such as images, videos, meshes, and animations. Existing strategies such as storing whole files for individual revisions or simple binary deltas, respectively consume significant storage and obscure semantic information. To overcome these limitations, in this paper we present a revision control system for digital images that stores revisions in form of graphs. Besides, being integrated with Git, our revision control system also facilitates artistic creation processes in common image editing and digital painting workflows. A preliminary user study demonstrates the usability of the proposed system.

## I. Introduction

In multimedia design and development, there is a wide range of contents such as text, images, video and audio that need to be created and edited. Recently, collaborative forms of multimedia development have revealed useful for authoring, editing, collecting, and producing digital content [1],[2],[3].

When the development of multimedia is carried out in collaborative and integrated design environments, revision control becomes essential to calculate, represent, and store differences between successive versions of the developed digital objects. Indeed, the development process can involve many authors with potentially different skills and different authoring tools. The standard paradigm of collaborative editing lies on sharing files between various instances of applications. This requires maintaining consistency of the versions and dealing with concurrent edits in the same part of a multimedia.

So far, revision control has been widely deployed for text files, while binary files have received relatively less attention. Very few methods are available for the efficient storage of modifications on any binary content. Existing strategies such as storing whole files for individual revisions or simple binary deltas could consume significant storage and obscure semantic information. This can be inconvenient for multimedia applications that use a significant amount of binary data, such as images, videos, and animations.

In this work, we present a revision control system for digital images that is intended to be used in a collaborative scenario of multimedia development. The core idea is the storage of graphical operations using a directed acyclic graph (DAG), where each vertex contains the result of a performed editing operation. This representation is suitable for nonlinear revision control and does not require storing the whole modified image.

## II. Related work

Revision control systems are widely used tools in software development primarily aimed on managing versions of software source code during implementation [4]. Since source code is written in plain text format, these systems are generally easily applicable to text documents. However these systems have traditionally lacked specific support for handling binary files. Even modern revision control systems like Git still offer limited built-in support for binary content [5]. Some extensions have been proposed (e.g., Git LFS),[1] but they are meant to add support for storing large binary files rather than for versioning.

Revision control systems can be distinguished as *state-* or *change-based* depending on the adopted history model, that is, how the history of changes to artifacts is stored [8]. Change-based systems (also known as operation-based) store the operations actually performed between two succeeding versions in the revision control system. Instead, state-based systems store the history of changes as revisions of the versioned artifacts as they existed at various points in time. Most existing general-purpose revision systems such as Git and SVN employ a state-based model. These modern revision control tools save storage space by computing and persisting only the difference (i.e., delta [7]) between succeeding revisions, while preserving the full state of a few special versions – like initial or final (head) revisions. However, when dealing with image data, these systems store separate images for each revision, thus wasting storage space. This issue alone hampers the adoption of revision control systems in managing digital images.

Besides, commonly used image-editing software tools offer a very limited control of image versions. For example, Adobe Photoshop provides a version history that retains the undone operations on a stack. Hence, the user can easily jump to any recent state of the image created during the current working session. However, high-level operations such as comparison (diff), branching, or merging of different versions are not available in Photoshop, nor in other current drawing tools.

To overcome these limitations, Chen et al. [6] proposed a solution for the nonlinear revision control of images, imple-mented as a plug-in of the image editor GIMP,[2] which tracks user editing actions in form of graphs to visualize revisions and support complex operations such as branching and merging.

---

[1] https://git-lfs.github.com

Along with the idea of the work proposed in [6], we have developed a revision control system for digital images that implements a hybrid approach combining the use of graphs with both state- and change-based revision control. First, like other change-based systems [8], we expose history as Directed Acyclic Graph (DAG, see Section III for more) to represent spatial, temporal, and semantic dependencies between successive recorded image-editing operations that are stored as graph nodes. Then, like state-based revision control system, we allow users to store important revisions as binary files, thanks to the integration with Git.

Furthermore, our system differs from the one proposed in [6] also in the following aspects:

- Rather than forking the GIMP image editing software, we developed our prototype almost from scratch, using Python and leveraging only stable and well-known libraries (e.g., Mathplotlib, Pillow). While the forked version of the tool developed by Chen et al. [6] is now obsolete (i.e., no one has ported the changes to the newer versions of GIMP), our approach promises to be more future-proof by allowing users to freely upgrade libraries as new updates are released.
- Our system is integrated with Git and GitHub, thus it enables teams to collaborate in the creation of both visual material and text artifacts. Besides, our system works well with the Git-LFS extension, thus allowing users to check out only the entire image files needed for the current task at hand, and just symbolic references for the other images not needed.
- Finally, instead of using custom formats [6], we store meta-data and project-related information using files in standard formats such as JSON and CSV.

## III. THE REVISION CONTROL SYSTEM

The core data structure of our system is a Direct Acyclic Graph (DAG) that is used to store the revisions as deltas. DAG nodes represent image editing operations with relevant information such as the type of operation and its parameters, the author who applied the operation, the time of the application and eventual notes. DAG edges represent the relationships between the operations. A (directed) sequential path between two nodes implies a spatial and/or semantic dependency between operations and the path direction gives information about their temporal order. Spatial dependency considers the spatial relationships between operations. Two operations are spatially independent if they are applied to non-overlapping regions. For example, *drawing a shape* and *coloring it* are spatially dependent operations. Conversely, *drawing a shape* and *coloring another existing shape* are independent operations. Semantically independent operations are rigid transformations (translation, rotation), deformation (scale, shear, perspective), color adjustment (hue, saturation, brightness, contrast, gamma) and filter (e.g., blur, sharpen).

[2]http://gimp.org

Table I
OPERATIONS LEADING TO THE CREATION OF A NEW NODE IN THE DAG.

| Type | Operation |
| --- | --- |
| *Rigid Transformation* | Mirror, Flip, Transpose |
| *Deformation* | Scale |
| *Color and Filter* | Histogram, Brightness, B&W, Sepia, Invert, Solarize, Posterize |
| *Edit* | Crop, Text, Reset |
| *Brush* | Brush |
| *Load image* | New, Import |

Multiple parallel paths between two nodes imply independent operation sequences, namely those that apply on disjoint regions. The DAG records the user editing operations and dynamically grows as more revisions are committed. Each revision in our system is a sub-graph of the DAG containing the root node which represents the act of initialization, i.e. opening an empty canvas or loading an existing image. The state of the revision is always equivalent to the result generated by traversing its corresponding sub-graph. It should be noted that in our system, the DAG encodes only actions, not whole images.

### A. Revision control commands

Based on the DAG data structure, our system provides the primary mechanisms for automatic resolving and merging multiple revisions with potential conflicts, as well as a user interface that allows manual change and intervention on automatically merged images. The implemented revision control commands include review, addition, branch, merge, and conflict resolving. We also provide an image diff tool that can be particularly useful to track not only low-level (pixel-based) differences but also high-level differences that modify the semantic content conveyed by the image. All these functionalities are offered through a friendly user interface. Table I lists the editing operations that lead to the creation of a new node in the DAG. In the following, we briefly describe the main revision control commands implemented in our system.

*1) Diff:* While the classic line-based `diff` command [7] is commonly used to extract differences between text files, there is no such well-defined difference tool for images. Among general image comparison visualization approaches, popular ones include side-by-side comparison (e.g. Adobe Bridge, Perforce), layer-based difference (e.g. PixelNovel), per-pixel difference, image overlay (e.g. Wet Paint [10]), and flickering difference regions (e.g. the compare utility of ImageMagick). These approaches are designed to handle only low level bitmap differences, with little information about the editing semantics. In contrast, following the idea of Chen et al. [6], our system realizes an informative diff by recording all the relevant high level information in a DAG. Two visual mechanisms for diff revision are available in our system. The first is the visualization of the DAG itself (fig. 5), which users can directly interact with to obtain visual clues about the involved editing operations. The second is a standalone diff UI that can be triggered by right-clicking on the DAG window. A new window is opened where the user is required to specify a source node and a target node. The diff UI provides a side-by-side comparison between revisions as well as sequential replay via a slider (fig. 1).

*2) Merge:* Unlike text, merging two images requires complex procedures in order to identify the difference between them. In text files changes are identified through line by line comparison. Instead, with binary files it is difficult to define which parts of an image have been changed. One way to implement the merge of two images is to represent images as a matrix whose elements contain information about each pixel within the image. In particular, rows ($x$) and columns ($y$) are the pixel coordinates, and each element of the matrix is an array of four elements, i.e., the three RGB (red, green, blue) values in $(0, 255)$ that determine the pixel color, plus the alpha level, which determines the opacity. Then, using this representation, applying the merge operator on two different DAG nodes will generate a new node including all changes (fig. 2). As such, the new node will be the obtained by applying the changes identified through the the pixel-wise diff between the two image matrices, that is, comparing the changes between the two arrays at the same position (coordinates). Should images modify the same pixel, rather than generating a conflict, we assume the 'latest' change to simply overwrite the other (we recall that all actions are recorded with a time-stamp).

*3) Commit:* Our prototype is capable of using Git to store images revisions locally as *commits*. Committing revisions is one of the most frequently used revision control commands. In our system, to save the current work progress as a new binary revision, users can simply issue the command `git commit` from the revision control window. Initially, when the repository is empty, the user adds the initial image and commits it as revision 0. Further commits replace individual nodes corresponding to delta changes, defining a new head revision. Although users can commit revisions whenever they like, it is generally unnecessary for them to do so in an action-wise fine-grained fashion, since our system can record all the actions and flexibly visualize them with DAG. As a general guideline, users proceed to commit revisions when one of the following two conditions is met: 1) some milestone of the work is achieved or 2) users would like to try out different variations. In the latter case, the committed revision can be used as a branch point for future reference or revision rollback.

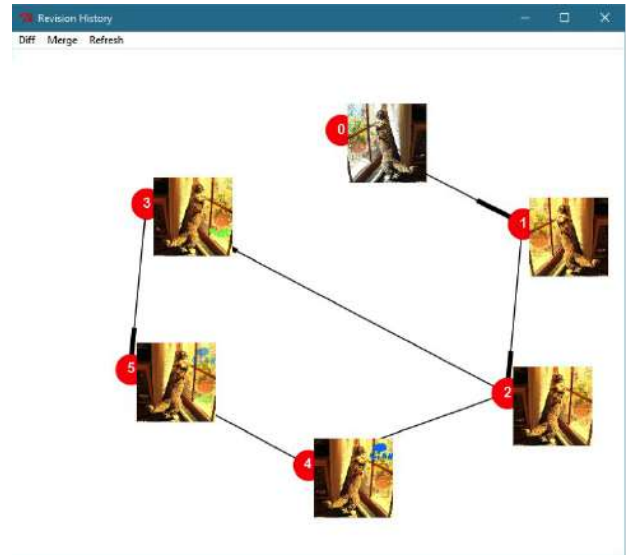

Figure 2. An example of two-way image merge: from node 2 a new branch (node 3) and revision (node 4) are started, which are later merged into the final revision (node 5).

*4) Push and Pull:* Our prototype is also capable of using Git repositories shared online on GitHub. When a new project is started, the URL of the associated GitHub repository must be provided, along with the path of a local folder where the content of the remote repository will be cloned locally. Then, it is possible to execute the `git push` and `git pull` commands, respectively, to store remotely pending commits and download updates pushed by other collaborators.

Furthermore, our system is compatible with the Git-LFS (Large-File Storage) extension that allows, when enabled, the storage of large files to a separate repository. The original repository will only contain pointers to the actual large files. Thus, user can decide to actually download these files only if need be. As such, the use of Git-LFS is recommended for speeding up the access to project repositories that host a number of very large files as in the case of images.

*B. Implementation*

The proposed system is entirely implemented in Python 2.7. The following libraries have been employed:

- `PIL` and `Pillow` for image editing;
- `Networkx` to create the structure of the DAG;
- `Mathplotlib` to visualize nodes and arcs of the DAG;
- `Json` to build the files necessary for handling the DAG;
- `Git-lfs` and `gitdb2` to implement support for Git and GitHub.

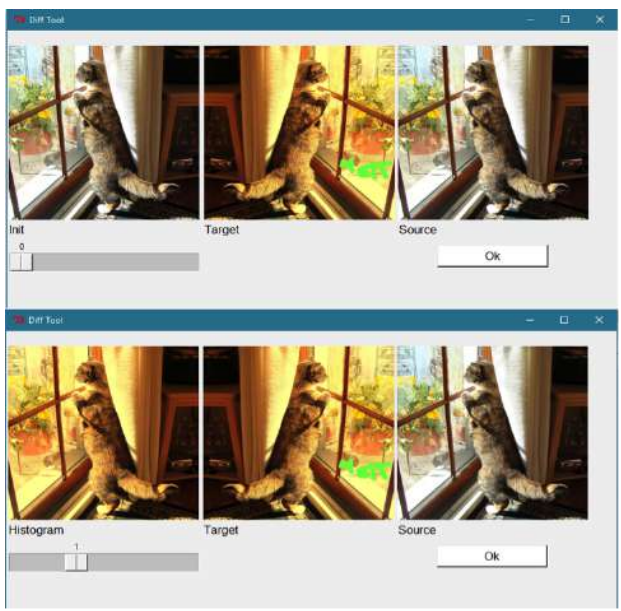

Figure 1. Diff UI. The preview window (left) shows the editing process between two revisions (Target and Source). Users can manually drag the slider for a particular state.

We have fully integrated our revision control system with an existing open source tool, called *Image Editor* [9], which we forked and evolved considerably. Specifically, we have modified the editor itself and added two main components: the *logger*, for recording user actions, and the *replayer*, for replaying actions starting from the stored deltas. The logger silently records user-editing actions in the background in the form of text logs that can be replayed in the image editor via the replayer. Precisely, when a user applies an operation that modifies the image, the logger detects the type of operation and creates three log files:

- a CSV file containing information related to a node;
- a JSON file describing the structure of the DAG;
- a CSV file containing the revision (delta).

The last file is used by the replayer to reconstruct the image after edition operation. All the user actions involving changes in the image are annotated in the log files to enable the porting of the project among different platforms and operating systems. In addition, for each new project the system creates the file `Project.properties` containing the project information (e.g author, revision format,...). All recorded logs and other revision control information are stored in a local repository (i.e. the project directory) and they can eventually be stored also in a repository shared online on GitHub. The logs are analyzed and transformed into DAG. When a new revision is committed, the corresponding action logs are transferred into the repository and recorded in the DAG.

### C. User Interface

The whole User Interface (UI) has been developed using the `Tkinter` library of Python. Below we describe the main usage scenarios of our system through the UI. When the user logs to the system, the starting window appears as depicted in fig. 3. As in most graphic software, there is a tool bar (on the left) with buttons to apply the different functions are available. On the top there is a menu bar where all the additional functionalities can be activated, such as project creation and saving. The canvas on the center of the window allows the user to operate on the image. In the orange bar on the bottom the user can add an annotation to the image. This feature is useful in collaborative settings to annotate the applied changes with a motivation. To start the work, it is necessary to create a new project. When the user selects the command `File>New Project` a pop-up window appears (fig. 4) that enables the user to enter the information related to the project, such as the name and the source image (if any). Users can also specify the author name, useful to keep track of *who changed what* in collaborative settings, and the file format to be used. Currently the system supports the following image formats: jpeg, png, tiff, and bmp.

By selecting Show History from either the toolbar or the menu, the DAG related to the open project will be displayed (fig. 5). Each node will be identified using both a number and a thumbnail. The number is useful to indicate the semantic relationship and the timeline of revisions. It is also possible to interact with each node: the right-click shows the details about the node (name, operator, time) and the left-click allows choosing among different options such as apply the `diff` and `merge` operators, and refresh the DAG in order to visualize the latest changes applied.

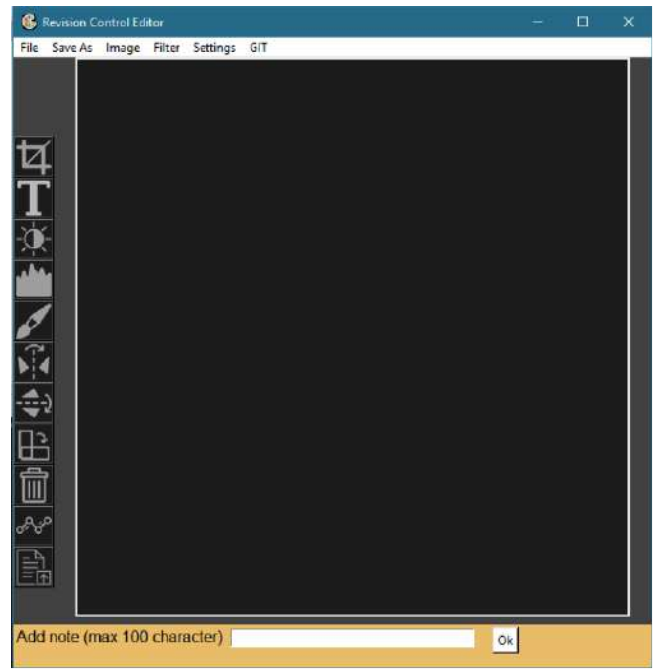

Figure 3. Starting window of the UI.

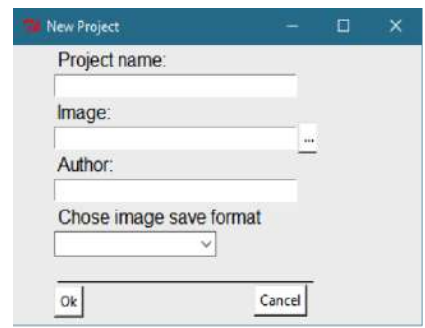

Figure 4. New Project pop-up window.

For each node, a double-click opens the Node Info window (fig. 6), which shows the revision image and other information such as the author, the image size, and its annotations. In the same window, users can add or edit existing annotations or even start a new revision branch from the current image (see fig. 6).

## IV. USER TEST

In order to evaluate the system, a usability test has been conducted. The adopted procedure is compliant to eGLU usability protocol used by the Italian Public Administration [11]. This protocol, which is one of the most used for simplified usability tests, it has been enriched by the eGLU-M for mobile systems and by the Usability Glossary of WikiPA project [11]. The protocol has been defined in order to be adaptable to different kinds of software. In this case, the following documents related to eGLU 2.1 protocol have been used:

- data of participants;
- description of the task;
- questionnaire for computing the Net Promoter Score (NPS);
- questionnaire for computing the System Usability Scale (SUS);
- table of results.

*A. Sample*

The participants involved in the study were 5 computer science students. As argued by Nielsen [12], 5 users would be expected to find 85% of the usability problems. The sample in our test was homogeneous since all the subjects had a high computer experience and no experience with the revision control system.

*B. Assigned tasks*

Each user was asked to complete the following tasks:
1) create a project;
2) apply three changes to the image (including resize and filtering operators);
3) add an annotation in the main window;
4) visualize the revision history;
5) visualize the information related to node 1 and to update the annotation;
6) add a branch to node 1;
7) apply the `diff` operator to visualize the changes made from node 0 to node 4,
8) `merge` two nodes;
9) to make a `pull` and a `push` from and to a GitHub project.

We observed that all the students took similar amounts of time for each task and anyone encountered some difficulties in using the system. In particular, 2 users had some problems in completing task #2 since they were not able to find the filtering operators (and hence they used the Brightness operator in place of a filtering operator). Moreover, 1 user encountered some difficulty in adding the annotation (task#3) and 4 users had difficulty in using Git (task #9). Indeed, task#9 was the most time-consuming task. However, this was somewhat expected, since none of the participants was familiar with Git commands and collaborative development.

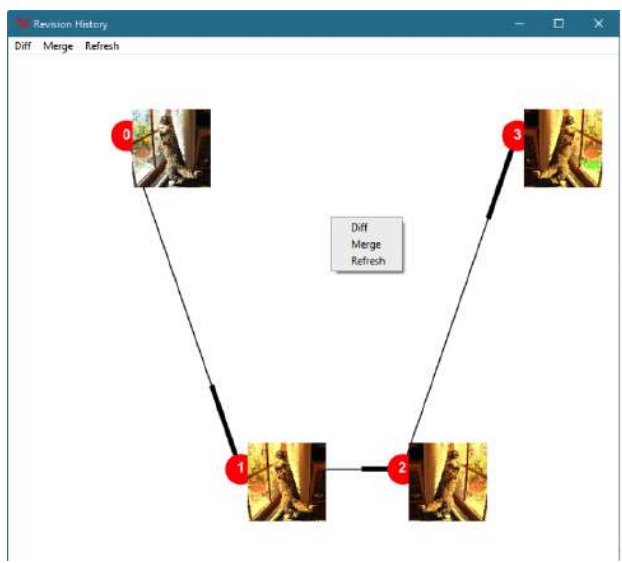

Figure 5. Example of DAG related to a project.

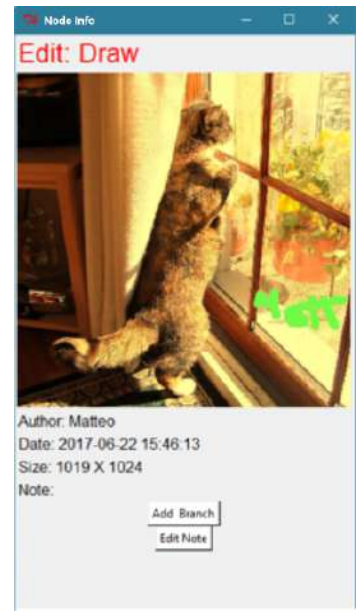

Figure 6. Node Info window.

*C. System Usability Scale*

The System Usability Scale (SUS) is a quick and reliable tool for measuring the usability of an artifact (software code or multimedia) [13] and is currently used to evaluate the usability of different kinds of products and services. The questionnaire consists of 10 items using a 5-point Likert scale. For each item the participant's response is converted in a new number and then it is summed to all the scores to obtain a number in [0,100]. The obtained score represents the average satisfaction level of the sample used in the user test. In this particular case, the sample is not very representative of the typical users of the system, thus the results can not be generalized, but they are indicative of possible usage problems. Based on literature, a SUS score above 68 can be considered above the average. In our case an average SUS score of 73.5 was achieved (fig. 7), hence we can conclude that at its primary stage the usability of the system is acceptable.

## V. CONCLUSIONS

In this work, we have presented a system for the revision control of digital images. Unlike other existing systems for image revision control, the proposed system adopts a hybrid approach that saves user editing actions as direct acyclic graphs (to save storage space), but also allows user to save important milestones revisions as binary files. Finally, thanks to the integration with the popular revision control system Git, our system is capable of supporting distributed collaborative team work. The usability study conducted with a few subjects provide initial evidence that our revision system is easy to use. However, the usability test focused on single-user tasks that capture the predominant usage scenarios in multimedia design. Further tests should be carried out in order to evaluate our system in collaborative scenarios of multimedia development.

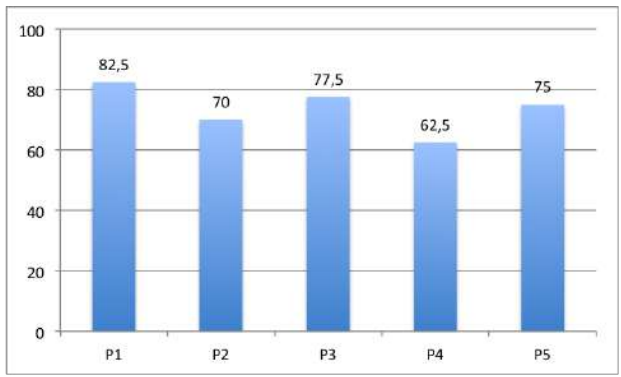

Figure 7. Results of the usability test on 5 participants (P1,P2,...,P5).

The image editor developed so far is intended as a proof of concept for our idea. For this reason, it offers only a minimal set of graphical operations. Nevertheless, it can be already used in real-world scenarios, e.g., by graphic designers. As a forthcoming future work, we intend to make our solution more appealing by integrating it within a more sophisticated and widely used open-source image editor.

## Acknowledgement

This work is partially funded by the project "Creative Cultural Collaboration" (C3) under the Apulian INNONETWORK programme, Italy.